\documentclass[a4]{article}
\usepackage{a4wide}

\usepackage{helvet}         
\usepackage{courier}        
\usepackage{type1cm}        

\usepackage{graphicx}        
\usepackage{multicol}        
\usepackage[bottom]{footmisc}

\usepackage{cite}

\usepackage{amsmath}
\usepackage{amssymb}
\allowdisplaybreaks[3]

\newcommand{\interior}[1]{\overset{\smash{\raisebox{-0.12ex}{$\scriptstyle\circ$}}}{#1}\rule{0pt}{2.3ex}}

\begin{document}

\title{High-order compact schemes for Black-Scholes basket options}
\author{Bertram D{\"u}ring\thanks{Department of Mathematics, University of
  Sussex, Pevensey II, Brighton, BN1 9QH, United Kingdom,
  Email:~b.during@sussex.ac.uk} \and Christof Heuer\thanks{Lehrstuhl f{\"u}r Angewandte Mathematik und
Numerische Analysis, Fachbereich C, Bergische Universit{\"a}t
Wuppertal, Gau{\ss}str. 20, 42119 Wuppertal, Germany, Email:~cheuer@uni-wuppertal.de}}
%
%
\maketitle

\abstract{We present a new high-order compact scheme for the
  multi-dimensional Black-Scholes model with application to European
  Put options on a basket of two underlying assets. The scheme is
  second-order accurate in time and fourth-order accurate in space. Numerical examples confirm that a standard
  second-order finite difference scheme is significantly
  outperformed. }

\section{Introduction}
The multidimensional Black-Scholes model for option pricing
(e.g. \cite{Wil98}) considers $n\in \mathbb{N}_{\geq 2}$ underlying
assets $S_i\in [0,\infty[$ for $i = 1, \ldots , n$,  where each asset follows a geometric Brownian motion,
\begin{align}\label{sde_for_stock}
{\rm d} S_i(t)& =\mu_i S_i(t){\rm d}t + \sigma_i S_i(t){\rm
  d}W^{(i)}(t),
\end{align}
where $\mu_i\in \mathbb{R}$ is the drift and $\sigma_i\geq 0$ is the
volatility of the asset $S_i$, respectively, for $i=1 , \ldots , n$
and ${\rm d}W^{(i)}(t)$ denotes a Wiener Process at time $t\in[0,T]$
for some $T>0$. The correlation between the assets is given by ${\rm
  d}W^{(i)}(t){\rm d}W^{(j)}(t)=\rho_{ij}{\rm d}t$. The Lemma of
It{\^o} and standard no-arbitrage arguments lead to the following
(backward in time) parabolic partial differential equation with mixed
second-order derivative terms for the
option price $V=V(S_1,S_2,\dots,S_n,t)$ (see, e.g.\ \cite{Wil98}),
\begin{align*}
\frac{\partial V}{\partial t}+ \frac{1}{2}\sum\limits_{i=1}^{n}\sigma_i^2 S_i^2 \frac{\partial^2 V}{\partial S_i^2} + \sum\limits_{\substack{i,j=1\\ i<j}}^{n}\rho_{ij}\sigma_i \sigma_j S_i S_j \frac{\partial^2 V}{\partial S_i \partial S_j} + \sum\limits_{i=1}^n rS_i \frac{\partial V}{\partial S_i} -rV  =&0,
\end{align*}
with $S_i>0$, $t\in [0,T[$ and $r\geq 0$ denoting the riskless interest rate.
When examining a European Put basket option, the final condition is given by
\begin{align*}
V(S_1, \ldots ,S_n,T)=&\max\biggl(K-\sum\limits_{i=1}^n \omega_i S_i ,0 \biggr) ,
\end{align*}
where the asset weights satisfy $\sum\limits_{i=1}^n \omega_i=1$ and additionally $\omega_i> 0$
for $i=1, \ldots,n$ if we have short-selling restrictions. Suitable
boundary conditions are discussed later. 

The transformations
\begin{align}\label{transformations}
 x_i =& \frac{\gamma}{\sigma_i}\ln\left(\frac{S_i}{K}\right), \quad \tau  =  T-t \quad \text{ and } \quad u=e^{r\tau}\frac{V}{K},
\end{align}
where $\gamma>0$ is a constant scaling parameter, yield the (forward
in time) parabolic partial differential equation
\begin{align}\label{usedpdemutliblsbasket}
u_{\tau} -\frac{\gamma^2}{2}\sum\limits_{i=1}^n \frac{\partial^2 u}{\partial x_i^2} - \gamma^2 \sum\limits_{\substack{i,j=1\\i<j}}^n \rho_{ij} \frac{\partial^2 u}{\partial x_i \partial x_j} + \gamma \sum\limits_{i=1}^n \left[ \frac{\sigma_i}{2}-\frac{r}{\sigma_i}\right]\frac{\partial u}{\partial x_i} = & 0,
\end{align}
where $x\in\mathbb{R}^n$ and $\tau\in \Omega_\tau=]0,T]$. Under the same transformations the initial condition for a European Put basket is given by
\begin{align}\label{general_Initial_Cond_Bls_Basket}
u(x_1, \ldots ,x_n,0)=&\max\biggl(1-\sum\limits_{i=1}^n \omega_i e^{\frac{\sigma_i x_i}{\gamma}} ,0 \biggr).
\end{align} 

When looking for numerical methods to approximate solutions
to problem \eqref{usedpdemutliblsbasket},
\eqref{general_Initial_Cond_Bls_Basket}, subject to suitable boundary
conditions, finite difference schemes can be employed, at least for
space dimensions up to three.
Standard discretisations, however, only yield second-order
convergence in terms of the spatial discretisation
parameter.
Alternatively, high-order compact schemes can be used which only use
points on a compact computational stencil, while having fourth-order
consistency in space, see for example
\cite{KarZha02,SpoCar01,TaGoBh08,DuFo12,DuFoHe14} and the references therein.
A drawback is that the derivation of high-order compact schemes (and
their numerical stability analysis) is algebraically demanding, hence
most works in this area restrict themselves to the one-dimensional
case. An additional complication is present in
\eqref{usedpdemutliblsbasket}  in form of the mixed second-order
derivative terms.

In a forthcoming paper \cite{DueHeu14} we derive new high-order compact schemes for a
rather general class of linear parabolic partial differential
equations with mixed second-order
derivative terms and time- and space-dependent coefficients in
arbitrary space dimension $n \in \mathbb{N}$. In the present paper we
focus on  the
  multi-dimensional Black-Scholes model \eqref{usedpdemutliblsbasket},
\eqref{general_Initial_Cond_Bls_Basket}. We present a new high-order compact scheme which is
  second-order accurate in time and fourth-order accurate in
  space. To ensure high-order convergence in the presence of the
initial condition \eqref{general_Initial_Cond_Bls_Basket} with low
regularity we employ the smoothing operators
of Kreiss et al.\ \cite{KrThWi70}.
Numerical examples for pricing European Put options on a basket of two underlying assets confirm that a standard
  second-order finite difference scheme is significantly
  outperformed.

\section{Discrete two-dimensional Black-Scholes equation}

For the discretisation of \eqref{usedpdemutliblsbasket} with $n=2$ we
replace the spatial domain by the rectangle $\Omega = [x_{\min}^{(1)}, x_{\min}^{(1)}]\times [x_{\min}^{(2)}, x_{\min}^{(2)}]$ with $-\infty<x_{\min}^{(i)}<x_{\min}^{(i)}<\infty$ for $i=1,2$. On $\Omega$, we define the grid
\begin{equation}\label{n_dimensional_Grid_general}
G^{(2)}_h = \big\{ (x^{(1)}_{i_1}, x^{(2)}_{i_2}) \in \Omega \text{ } | \text{ } x^{(k)}_{i_k} = x_{\min}^{(k)} +i_k h,\,
 1\leq i_k \leq N_k,\,k=1, 2 \big\},
\end{equation}
where $h>0$, $N_k \in \mathbb{N} $ and $ x_{\max}^{(k)} =
x_{\min}^{(k)} +N_k h$ for $k=1,2$. By $\interior{G}^{(2)}_h$ we
denote the interior of $G^{(2)}_h$. We present the coefficients of a
semi-discrete scheme of the form
\begin{align*}
\sum\limits_{j_1=i_1-1}^{i_1+1} \sum\limits_{j_2=i_2-1}^{i_2+1} \left[\hat{M}_{j_1,j_2} \partial_{\tau}U_{j_1,j_2}(\tau) + \hat{K}_{j_1,j_2} U_{j_1, j_2}(\tau) \right] = & \tilde{g}(x,\tau) ,
\end{align*}
at time $\tau$ for each point $x\in \interior{G}_h^{(2)}$
for the two-dimensional Black-Scholes equation using $n=2$ in
\eqref{usedpdemutliblsbasket}. By $U_{j_1,j_2}(\tau)$ we denote the
approximation of $u(x_{i_1}^{(1)},x_{i_2}^{(2)},\tau)$ after
semi-discretisation in space with $\bigl(
  x_{i_1}^{(1)},x_{i_2}^{(2)}\bigr) \in G_h^{(2)}$. 

The general idea underlying the derivation of the high-order compact
scheme is to operate on the differential equation \eqref{usedpdemutliblsbasket} as an
additional relation to obtain finite difference approximations for
high-order derivatives in the truncation error. Inclusion of these
expressions in a central difference method for equation \eqref{usedpdemutliblsbasket}
increases the order of accuracy to fourth order while retaining a compact stencil.
A detailed derivation of this scheme and a thorough von Neumann
stability analysis are
presented in a forthcoming paper \cite{DueHeu14}. In the
two-dimensional case we obtain the following coefficients 
\begin{align*}
\hat{K}_{i_1,i_2}  = & -\,{\frac {2{\gamma}^{2}{\rho^{2}_{{12}}}}{3{h}^{2}}}
+\,{\frac {5{\gamma}^{2}}{3{h}^{2}}}
+\frac{ \left( \frac{\sigma_{{1}}}{2}-{\frac {r}{\sigma_{{1}}}} \right) ^{2}}{3}
+\frac{ \left( \frac{\sigma_{{2}}}{2}-{\frac {r}{\sigma_{{2}}}} \right) ^{2}}{3},
\\
\hat{K}_{i_1\pm 1,i_2}  = & \,{\frac {{\gamma}^{2}{\rho^{2}_{{12}}}}{3{h}^{2}}}
\pm \frac{\gamma\, \left( \frac{\sigma_{{1}}}{2}-{\frac {r}{\sigma_
{{1}}}} \right) }{3h}
\mp \frac{\gamma\, \left( \frac{\sigma_{{2}}}{2}-{\frac {r}{\sigma_{{2}}}} \right) \rho_{{12}}}{3h}
- \frac{ \left( \frac{\sigma_{{1}}}{2}-{\frac {r}{\sigma_{{1}}}} \right) ^{2}}{6}
- \,{\frac {{\gamma}^{2}}{3{h}^{2}}},
\\
\hat{K}_{i_1,i_2\pm 1} = & \,{\frac {{\gamma}^{2}{\rho^{2}_{{12}}}}{3{h}^{2}}}
\pm \frac{\gamma\, \left( \frac{\sigma_{{2}}}{2}-{\frac {r}{\sigma_{{2}}}} \right)} {3h}
\mp \frac{\gamma\left( \frac{\sigma_{{1}}}{2}-{\frac {r}{\sigma_{{1}}}} \right)\,\rho_{{12}} } {3h}
- \frac{ \left( \frac{\sigma_{{2}}}{2}-{\frac {r}{\sigma_{{2}}}} \right) ^{2}}{6}
- \,{\frac {{\gamma}^{2}}{3{h}^{2}}},
\\
\hat{K}_{i_1\pm 1,i_2-1}  = & \pm \frac{ \left( \frac{\sigma_{{2}}}{2}-{\frac {r}{\sigma_{{2}}}}
 \right)  \left( \frac{\sigma_{{1}}}{2}-{\frac {r}{\sigma_{{1}
}}} \right) }{12}
- \frac{\gamma\, \left( \frac{\sigma_{{2}}}{2}-{\frac {r}{\sigma_{{2}}}} \right) }{12h}
\pm \frac{\gamma\, \left( \frac{\sigma_{{1}}}{2}-{\frac {r}{\sigma_{{1}}}} \right) }{12h}\\
& 
- \frac{\gamma\left( \frac{\sigma_{{1}}}{2}-{\frac {r}{\sigma_{{1}}}} \right)\,\rho_{{12}} }{6h}
\pm \frac{\gamma\, \left( \frac{\sigma_{{2}}}{2}-{\frac {r}{\sigma_{{2}}}} \right) \rho_{{12}}}{6h}
- \,{\frac {{\gamma}^{2}}{12{h}^{2}}}
\pm \,{\frac {{\gamma}^{2}\rho_{{12}}}{4{h}^{2}}}
- \,{\frac {{\gamma}^{2}{\rho^{2}_{{12}}}}{6{h}^{2}}} ,
\\
\hat{K}_{i_1\pm 1,i_2+1}  = &  \frac{\gamma\, \left( \frac{\sigma_{{2}}}{2}-{\frac {r}{\sigma_{{2}}}} \right) }{12h}
\mp \frac{ \left( \frac{\sigma_{{2}}}{2}-{\frac {r}{\sigma_{{2}}}
} \right)  \left( \frac{\sigma_{{1}}}{2}-{\frac {r}{\sigma_{{1
}}}} \right) }{12}
\pm \frac{\gamma\, \left( \frac{\sigma_{{1}}}{2}-{\frac {r}{\sigma_{{1}}}} \right) }{12h}\\
& 
+ \frac{\gamma\,\rho_{{12}} \left( \frac{\sigma_{{1}}}{2}-{\frac {r}{\sigma_{{1}}}} \right) }{6h}
\pm \frac{\gamma\, \left( \frac{\sigma_{{2}}}{2}-{\frac {r}{\sigma_{{2}}}} \right) \rho_{{12}}}{6h}
- \,{\frac {{\gamma}^{2}}{12{h}^{2}}}
\mp \,{\frac {{\gamma}^{2}\rho_{{12}}}{4{h}^{2}}}
- \,{\frac {{\gamma}^{2}{\rho^{2}_{{12}}}}{6{h}^{2}}},
\end{align*}
as well as
\begin{align*}
M_{i_1+1, i_2\pm 1} = & M_{i_1-1, i_2\mp 1}  = \pm \frac{\rho_{{12}}}{24} , & M_{i_1,i_2}  =& \frac{2}{3} , \\
M_{i_1 \pm 1,i_2}  = & \frac{1}{12}
\mp \frac{h \left( \frac{\sigma_{{1}}}{2}-{\frac {r}{\sigma_{{1}}}} \right) }{12\gamma} , & M_{i_1,i_2\pm1}  = & \frac{1}{12}
\mp \frac{h \left( \frac{\sigma_{{2}}}{2}-{\frac {r}{\sigma_{
{2}}}} \right) }{12\gamma} .
\end{align*}
Additionally, $\tilde{g}(x,\tau) = 0$ for $x\in \interior{G}_h^{(2)}$
and $\tau \in \Omega_{\tau}$. 
After presenting the high-order compact discretisation for the spatial interior we now discuss the boundary conditions. 

\section{Discretisation of the boundary conditions}\label{Lower_boundaries_n_dim_BLS_basket}
The first boundary we discuss is $S_i=0$ for some $i\in \{1, 2\}$ at time $t\in [0,T[$. Once the value of the asset is zero, it stays constant over time, see \eqref{sde_for_stock}. If only one asset reaches its minimum value, using $S_i=0$ for $i\in \{1,2\}$ in the multi-dimensional Black-Scholes equation with $n=2$ leads to the one-dimensional Black-Scholes equation for the asset $S_j$ with $j=\{1,2\}\setminus {i}$. One can either transform the solution of the one-dimensional Black-Scholes partial differential equation using \eqref{transformations} or derive a fourth-order compact scheme for these boundaries similarly to the space interior. If both asset values are minimal, we have
\begin{eqnarray}
\notag u(x_{\min}^{(1)},x_{\min}^{(2)},\tau)=u(x_{\min}^{(1)},x_{\min}^{(2)},0)
\end{eqnarray} 
for $\tau \in ]0, \tau_{\max}]$ after transforming with \eqref{transformations}.

Upper boundaries are boundaries with $S_i=S_i^{\max}>0$ with $i \in \left\{1,2\right\}$ at time $t\in [0,T[$. For a sufficiently large $S_i^{\max}$, we can approximate
\begin{align}\label{condition_upper_boundary}
\frac{\partial V\left(S_1, S_2, t \right) }{\partial S_i}\Big|_{S_i = S_i^{\max}} \equiv& 0,
\end{align}
with $S_k \in \left[S_k^{\min} , S_k^{\max}\right]$ for $k = \{1, 2\} \setminus \{i\}$. If only one underlying asset $S_i$ reaches its maximum value, using \eqref{condition_upper_boundary} in the two-dimensional Black-Scholes differential equation leads to the one-dimensional Black-Scholes differential equation for the underlying asset $S_j$ with $j = \{1,2\} \setminus \{i\}$. One can either transform the solution of this equation using \eqref{transformations} or transform the one-dimensional Black-Scholes differential equation using \eqref{transformations} and derive a fourth-order compact scheme for these boundaries. When both underlying assets reach their maximum value, we have
\begin{align*}
u(x_1^{\max},x_2^{\max}, \tau)  = & u(x_1^{\max},x_2^{\max}, 0)
\end{align*}
for $\tau \in ]0, \tau_{\max}]$ after using the transformations \eqref{transformations}.
Since the boundaries behave similar, we have
\begin{align*}
u(x_1^{\min},x_2^{\max}, \tau)  = & u(x_1^{\min},x_2^{\max}, 0),&
u(x_1^{\max},x_2^{\min}, \tau)  = & u(x_1^{\max},x_2^{\min}, 0),
\end{align*}
for $\tau \in ]0, \tau_{\max}]$.

\section{Time discretisation}
We use an equidistant time grid of the form $\tau = k\,\Delta \tau$ for $k = 0 , \ldots , N_{\tau}$ with $N_{\tau} \in \mathbb{N}$. Using a Crank-Nicolson-type time discretisation with step size $\Delta\tau$ leads to
\begin{align*}
&\sum\limits_{j_1=i_1-1}^{i_1+1} \sum\limits_{j_2=i_2-1}^{i_2+1} \left[ \hat{M}_{j_1,j_2} + \frac{\Delta \tau}{2}\hat{K}_{j_1,j_2}\right] U_{j_1,j_2}^{k+1}  \\
=& \sum\limits_{j_1=i_1-1}^{i_1+1} \sum\limits_{j_2=i_2-1}^{i_2+1} \left[ \hat{M}_{j_1,j_2}  - \frac{\Delta \tau}{2}\hat{K}_{j_1,j_2}\right]U_{j_1,j_2}^k + (\Delta \tau)g(x)
\end{align*}
at each point $\bigl(x_{i_1}^{(1)},x_{i_1}^{(2)}\bigr)\in G_h^{(2)}$, where only points of the compact stencil are used. By $U_{i_1,i_2}^k$ we denote the approximation of $u(x_{i_1}^{(1)},x_{i_2}^{(2)},\tau_k) $. For the Crank-Nicolson type time discretisation this compact scheme has consistency
order two in time and four in space. Thus, using $\Delta \tau \in\mathcal{O}\left(h^2\right)$, leads to fourth-order consistency in terms of the spatial stepsize $h>0$.

\section{Numerical experiments}
In this section we present numerical experiments for the Black-Scholes
European Puts basket option in space dimension $n=2$. According to
\cite{KrThWi70}, we cannot expect fourth-order convergence if the
initial condition $u_0$ is only in $C^0\left(\Omega\right)$. In
\cite{KrThWi70} suitable smoothing operators are identified in Fourier
space. Since the order of consistency of our high-order compact
schemes is four, we use the smoothing operator $\Phi_4$ (see \cite{KrThWi70}), given by its Fourier transformation
$$
\hat{\Phi}_4 (\omega)= \Biggl(\frac{\sin\bigl( \frac{\omega}{2}\bigr) }{\frac{\omega}{2}}\Biggr)^4 \left[ 1 + \frac{2}{3} \sin^2\left( \frac{\omega}{2} \right)\right].
$$
This leads to the smoothed initial condition given by
$$
\tilde{u}_0\left(x_1,x_2\right) = \frac{1}{h^2} \int\limits_{-3h}^{3h} \int\limits_{-3h}^{3h}\Phi_4 \left(\frac{x}{h}\right)\Phi_4 \left(\frac{y}{h}\right) u_0\left(x_1-x,x_2-y\right) \text{d}x \text{d}y,
$$
for any stepsize $h>0$, where $\Phi_4(x)$ denotes the Fourier inverse of
$\hat{\Phi}_4(\omega)$. If $u_0$ is smooth enough
in the integrated region around $\left(x_1, x_2\right)\in \Omega$, we
have $\tilde{u}_0 \left(x_1, x_2\right) = u_0\left(x_1,
  x_2\right)$. Thus it is possible to identify the points where
smoothing is necessary for a given initial condition.
This approach reduces the
necessary computations significantly. Note that as $h
\rightarrow 0$, the smoothed initial condition $\tilde{u}_0$ converges
to the original initial condition $u_0$ given in
\eqref{general_Initial_Cond_Bls_Basket}. Hence the approximation of
the smoothed problem tends towards the true solution of
\eqref{usedpdemutliblsbasket}. 

For examining the numerical convergence rate we use the relative $l^2$-error $\Vert U_{\text{ref}} -
U\Vert_{l^2}/\Vert U_{\text{ref}}\Vert_{l^2}$, as well as the
$l^{\infty}$-error $\Vert U_{\text{ref}} - U\Vert_{l^{\infty}}$, where $U_{\text{ref}}$
denotes a reference solution on a fine grid and $U$ is the
approximation. 
We determine the numerical convergence order of the schemes as the slope of the linear least square fit of the
individual error points in the loglog-plots of error versus number of
discretisation points per spatial direction. We compare the
high-order compact scheme to a standard second-order scheme, which results from
applying the standard central difference operators directly in
\eqref{usedpdemutliblsbasket} with $n=2$. We use the following parameters,
$$ \sigma_1 = 0.25, \; \sigma_2 = 0.35,\;\gamma = .25, \quad r = \text{log}(1.05), \;\omega_1 = 0.35 = 1-\omega_2.$$
and $K=10$. We set the parabolic mesh ratio $\Delta\tau/h^2 =
0.4$, but emphasise that neither the von
Neumann stability analysis presented in \cite{DueHeu14} nor additional numerical experiments reveal
any restrictions on this relation, indicating unconditional stability
of the scheme. We use different values
$\rho_{12} = -0.8,$ $\rho_{12} = 0$ and $\rho_{12} = 0.8$ for the correlation.
\begin{figure}[h]
      \includegraphics[width=5.5cm,height=4.5cm]{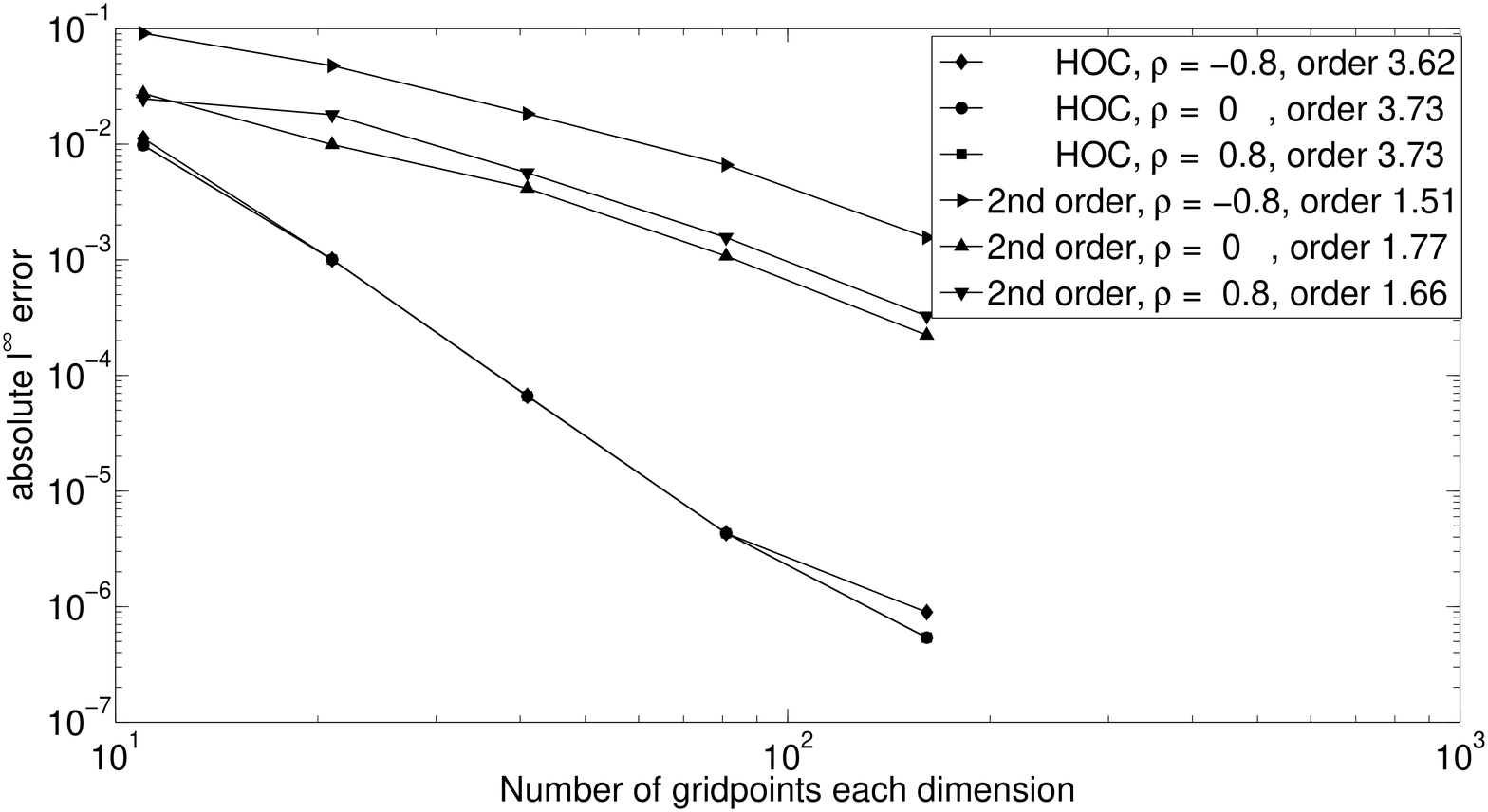}%
      \includegraphics[width=5.5cm,height=4.5cm]{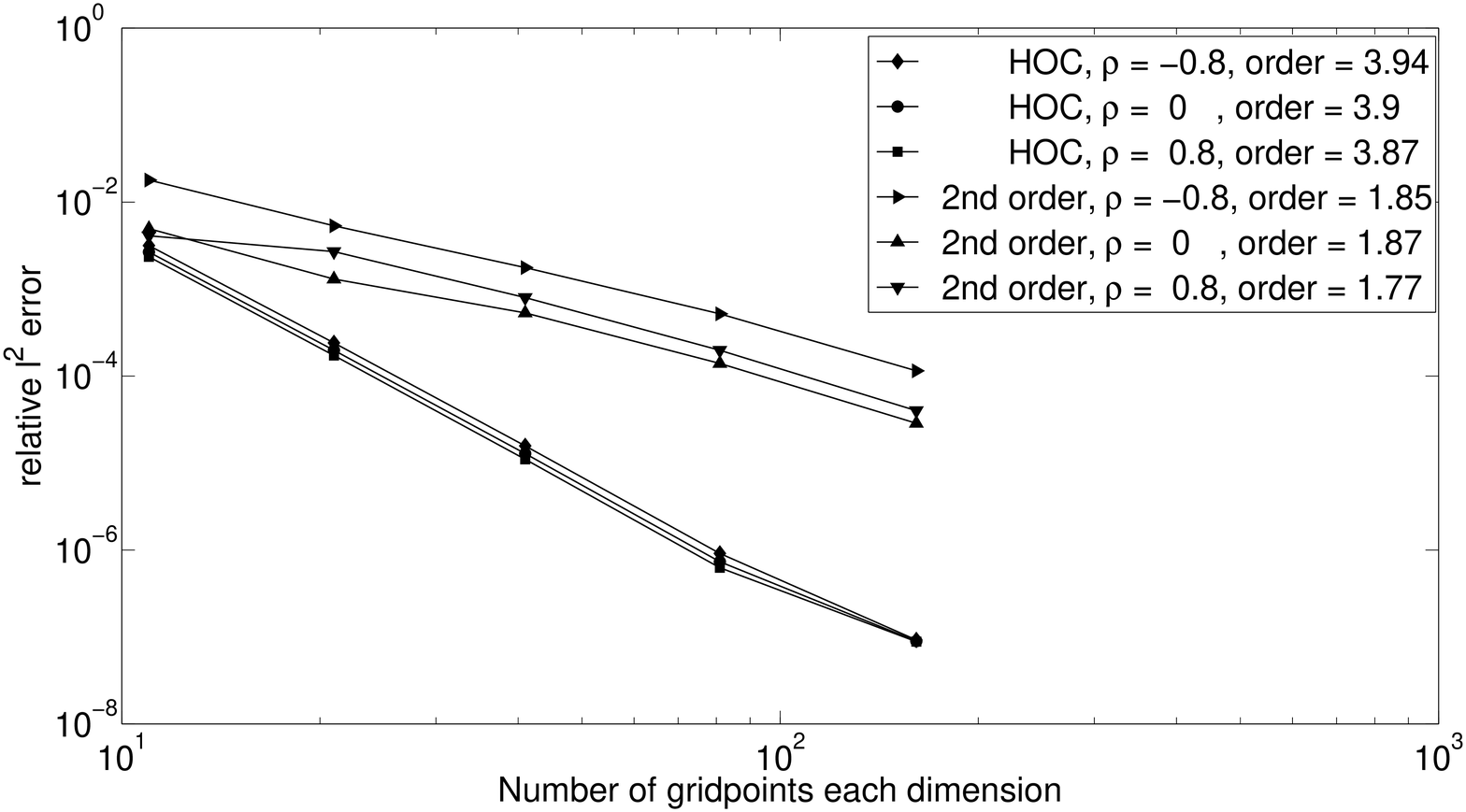}
      \caption{Absolute $l^{\infty}$-error and relative $l^2$-error for two-dimensional Black-Scholes Basket Put with smoothed initial condition.}
      \label{fig:loglog_plot}
\end{figure} 
In Fig.~\ref{fig:loglog_plot} we show plots of
the $l^{\infty}$-error and the relative $l^2$-error. 
The high-order compact scheme performs highly similar for the
three different correlation values, the points are almost
identical. The numerical convergence
orders for the high-order compact scheme range between  $3.62$ and
$3.73$ for the $l^{\infty}$-error, and between $3.87$ and $3.94$ for
the relative $l^2$-error.
The high-order compact
scheme significantly outperforms the standard second-order
discretisation in all cases.
\section*{Acknowledgement}
The second author was partially supported by the European Union in the
FP7-PEOPLE-2012-ITN Program under Grant Agreement Number 304617 (FP7
Marie Curie Action, Project Multi-ITN {\em STRIKE –- Novel Methods in Computational Finance}).

%
%
\bibliographystyle{spmpsci} 
\bibliography{Masterfile}  

\end{document}